\documentclass{article}
\usepackage[a4paper,margin=1in]{geometry}
\usepackage{amsmath,amssymb}
\usepackage{graphicx}
\usepackage{bm}
\usepackage{siunitx}
\usepackage{hyperref}
\usepackage{xcolor}
\usepackage{cite}
\usepackage{setspace}

\title{\bf Mechanism-driven CO$_2$ Capture and Activation on Two-dimensional Transition-metal Diborides}

\author{
Jakkapat Seeyangnok$^{1a}$,
Rungkiat Nganglumpoon$^{2}$,\\
Joongjai Panpranot$^{2b}$,
Udomsilp Pinsook$^{1b}$\\[1ex]
$^{1}$Department of Physics, Faculty of Science, \\
Chulalongkorn University, Bangkok, Thailand\\[1ex]
\texttt{$^{a}$jakkapatjtp@gmail.com}\\
\texttt{$^{b}$joongjai.p@chula.ac.th}\\
\texttt{$^{c}$Udomsilp.P@chula.ac.th}\\
}

\date{} 

\begin{document}

\maketitle

\begin{abstract}
The urgent need to mitigate rising atmospheric CO$_2$ levels motivates the search for stable, efficient, and tunable adsorbent materials. In this study, we employ first-principles density functional theory to investigate the adsorption of CO$_2$ molecules on two-dimensional hexagonal transition-metal diboride monolayers, M$_2$B$_2$ (M = Sc, Y, Ti, Zr, Nb). The adsorption energies, structural distortions, and bonding characteristics are systematically analyzed to understand how the metal center governs CO$_2$ activation. The calculated adsorption energies range from $-1.84$ to $-2.16$ eV (or $-1.98$ to $-4.42$), with Ti$_2$B$_2$ and Sc$_2$B$_2$ exhibiting the strongest CO$_2$ binding, while Y$_2$B$_2$, Zr$_2$B$_2$, and Nb$_2$B$_2$ show moderately strong chemisorption. Adsorption induces significant molecular activation, evidenced by elongated C–O bonds (1.27–1.29~\AA) and bent O–C–O angles (129–132$^\circ$), compared to the linear gas-phase configuration (1.17~\AA, 180$^\circ$). Charge analysis further reveals substantial electron transfer from the monolayer to CO$_2^{\delta-}$, consistent with strong chemisorption and structural deformation. Correspondingly, the shift toward less negative IpCOHP$(E_f)$ values indicates a pronounced weakening of the internal C–O bonds, reflecting increased population of antibonding $\pi^\ast$ orbitals. Ab initio molecular dynamics simulations show that the activated CO$_2^{\delta-}$ species is thermally sensitive: while most M$_2$B$_2$ surfaces retain stable adsorption at 300~K, Ti$_2$B$_2$ drives spontaneous CO$_2$ dissociation into CO and O, revealing a temperature-assisted activation pathway. These findings highlight how the choice of transition metal tunes electronic interactions, adsorption energetics, and activation pathways on M$_2$B$_2$ surfaces. Overall, this work identifies two-dimensional transition-metal diborides as promising candidates for next-generation CO$_2$ capture and activation technologies.
\end{abstract}


\section{Introduction}
    Global temperature records extending back to the late 19th century reveal a persistent warming trend. NASA’s Goddard Institute for Space Studies (GISS) recognized this increase as early as the 1980s, and their most recent analyses report a global temperature rise of more than 0.8~°C since 1951~\cite{hansen2010global,GISS_GISTEMP_2025}. Over the past century, human activities have driven a substantial increase in the emission of greenhouse gases, with carbon dioxide identified as the most significant contributor. Atmospheric CO$_2$ concentrations continue to rise steadily each year, increasing from 313~ppm in 1960 to more than 425~ppm in 2025~\cite{NOAA_CO2_Trends_2025}. This rapid escalation in CO$_2$ levels has become one of the most pressing environmental challenges of our time. The efficient capture and conversion of greenhouse gases, particularly CO$_2$, are therefore central to global climate-mitigation strategies and ecosystem protection efforts. Developing stable, efficient materials for CO$_2$ capture~\cite{li2017enhanced,tawfik2015multiple} remains a major scientific and technological challenge, as practical deployment requires a combination of high adsorption capacity, selectivity, and long-term chemical robustness. Despite significant advances in adsorption-, absorption-, and membrane-based separation technologies~\cite{yu2012review,saenz2023evaluating}, many available CO$_2$ sorbents still face limitations such as low selectivity, high regeneration costs, degradation under working conditions, or insufficient binding strength. These persistent drawbacks emphasize the ongoing need for new classes of materials with improved CO$_2$ affinity, tunable surface reactivity, and long-term stability under realistic industrial and environmental conditions~\cite{sanz2016direct,boot2014carbon}.

Two-dimensional (2D) materials have emerged as promising platforms for gas capture~\cite{yang2017gas,prakash2024recent,zhou2024advances} and surface-driven chemical processes owing to their large surface areas, exposed active sites, and tunable electronic properties. Considerable progress has been made on systems such as transition-metal dichalcogenides (TMDs), whose structural, electronic, and chemical properties have been widely reviewed~\cite{manzeli20172d,joseph2023review}, and which also exhibit highly reactive hydrogenated surfaces~\cite{lu2017janus,seeyangnok2024superconductivity,seeyangnok2024superconductivitywseh,qiao2024prediction,seeyangnok2025competition,sukserm2025half} including Li functionalization~\cite{xie2024strong,moseliseeyangnok}. These characteristics have enabled TMDs to demonstrate strong potential in gas capture and sensing applications~\cite{mirzaei2024resistive,llobet2024transition}. MXenes, another major family of 2D materials, have similarly been extensively investigated for their surface chemistry and functional versatility~\cite{li2022mxene,akhter2023mxenes}. Their surfaces also undergo hydrogenation~\cite{hm2x_jpcs,Seeyangnok2025phase_jap}, halogen functionalization~\cite{jsee_mo2cx2} and numerous studies have demonstrated their suitability for CO$_2$ capture and separation~\cite{mirzaei2024resistive,serafin2025mxenes,morales2021carbon,wang2023two,ozkan2025scaling}. Hexagonal boron nitride (h-BN), known for its chemical stability and tunable adsorption characteristics~\cite{zhang2017two}, has likewise shown promising gas adsorption behavior~\cite{guo2015co2}. More recently, boron-based 2D analogues such as BXenes have been predicted and reviewed as emerging materials with unique bonding characteristics~\cite{ramezanzadeh20252d,khan20242d}, and initial investigations indicate their potential for CO$_2$ capture~\cite{iravani2025environmental}.

These diverse families of 2D materials exhibit rich surface chemistry and offer valuable opportunities for designing efficient CO$_2$ adsorption and activation pathways at the atomic scale. Within this expanding landscape, 2D transition-metal borides have gained increasing attention as a new class of ultrathin materials characterized by robust structural stability and highly reactive surfaces~\cite{khan20242d,mir2022efficient}. In particular, hexagonal M$_2$B$_2$ monolayers have emerged as promising candidates for a wide range of applications. Several studies have explored their potential for metal-ion battery technologies~\cite{yuan2019monolayer,he2021computational,gao2021two,bo2018hexagonal,lu2024functionalizing,dai2025theoretical}, highlighting their favorable electronic and structural characteristics. Moreover, many hexagonal M$_2$B$_2$ BXene monolayers exhibit exceptionally active surfaces, as demonstrated by their strong tendency toward surface hydrogenation~\cite{han2023theoretical,han2023high,seeyangnok2025high_npj2d}, which has also been linked to enhanced superconducting behavior~\cite{han2023high,seeyangnok2025high_npj2d}. This pronounced surface reactivity suggests that these materials may also serve as efficient platforms for CO$_2$ capture, activation, and conversion. Despite these promising attributes, however, the adsorption behavior and surface chemistry of CO$_2$ on hexagonal M$_2$B$_2$ BXenes remain largely unexplored.

Given the global need for effective strategies to mitigate CO$_2$ emissions, identifying new materials capable of stable and selective CO$_2$ adsorption is of significant interest. In this work, we use first-principles calculations to investigate the structural, electronic, and surface chemical properties of hexagonal M$_2$B$_2$ (M = Sc, Y, Ti, Zr, Nb) monolayers and their interactions with CO$_2$. We first evaluate the geometry, phase stability, formation energies, and mechanical and dynamical robustness of the pristine monolayers to establish their viability as free-standing 2D materials. We then examine CO$_2$ adsorption by determining the most favorable binding configurations and analyzing the associated adsorption energetics and molecular distortions. To elucidate the adsorption mechanism, we investigate charge redistribution and bonding characteristics using Löwdin and Bader charge analysis and Crystal Orbital Hamilton Population (COHP) calculations. These results provide fundamental insight into the surface reactivity of this emerging class of 2D transition-metal borides and underscore their potential for CO$_2$ capture and related catalytic applications.

\section{Methods}
The structural and electronic properties of the M$_2$B$_2$ monolayers (M = Sc, Y, Ti, Zr, Nb) were investigated using density functional theory (DFT) as implemented in \textsc{Quantum~Espresso}~\cite{giannozzi2009quantum, giannozzi2017advanced}. Initial structures were generated in \textsc{VESTA}~\cite{momma2011vesta} and optimized with the BFGS quasi-Newton scheme~\cite{BFGS,liu1989limited} until the residual atomic forces were below $10^{-5}$~eV/\AA. A vacuum spacing of $20$~\AA\ was applied along the out-of-plane direction to avoid interactions between periodic images. Exchange–correlation effects were described using the GGA-PBE functional~\cite{perdew1996generalized} and PAW pseudopotentials~\cite{kresse1999ultrasoft}. Plane-wave cutoffs of 60~Ry (wavefunctions) and 240~Ry (charge density) were used. Brillouin-zone sampling employed a $21\times21\times1$ Monkhorst--Pack grid~\cite{monkhorst1976special} with a first-order Methfessel--Paxton smearing of 0.02~Ry~\cite{kresse1999ultrasoft}.

For the CO$_2$ adsorption calculations, a $3\times3\times1$ supercell of the M$_2$B$_2$ monolayers was employed together with a $7\times7\times1$ $k$-point mesh to ensure convergence of total energies and adsorption geometries. Long-range dispersion interactions, which play a non-negligible role in molecule–surface binding, were accounted for using Grimme’s D3 correction~\cite{grimme2010consistent}. To further investigate the bonding characteristics and surface chemical reactivity toward CO$_2$, Crystal Orbital Hamilton Population (COHP) analyses were performed using the \textit{LOBSTER} code~\cite{deringer2011crystal, maintz2013analytic, maintz2016lobster}.


The interaction strength between CO$_2$ and the $\mathrm{M_2B_2}$ ($\mathrm{M = Sc, Y, Ti, Zr, Nb}$) monolayers was assessed through the adsorption energy, $E_{\text{absorp}}$, which quantifies the energetic preference for CO$_2$ binding. It is expressed as

\begin{equation}
E_{\text{absorp}} = E_{\mathrm{M_2B_2+CO_2}} - \left( E_{\mathrm{M_2B_2}} + E_{\mathrm{CO_2}} \right),
\end{equation}

where $E_{\mathrm{M_2B_2+CO_2}}$ denotes the total energy of the relaxed CO$_2$--$\mathrm{M_2B_2}$ complex, $E_{\mathrm{M_2B_2}}$ is the energy of the isolated monolayer, and $E_{\mathrm{CO_2}}$ corresponds to the energy of a free CO$_2$ molecule calculated in vacuum. A negative value of $E_{\text{absorp}}$ signifies that the adsorption process is exothermic and thus thermodynamically favorable.

\section{Results and discussion}
\subsection{Crystal structure and stability of hexagonal M$_2$B$_2$ monolayers}
\begin{figure}[h!]
    \centering
    \includegraphics[width=10cm]{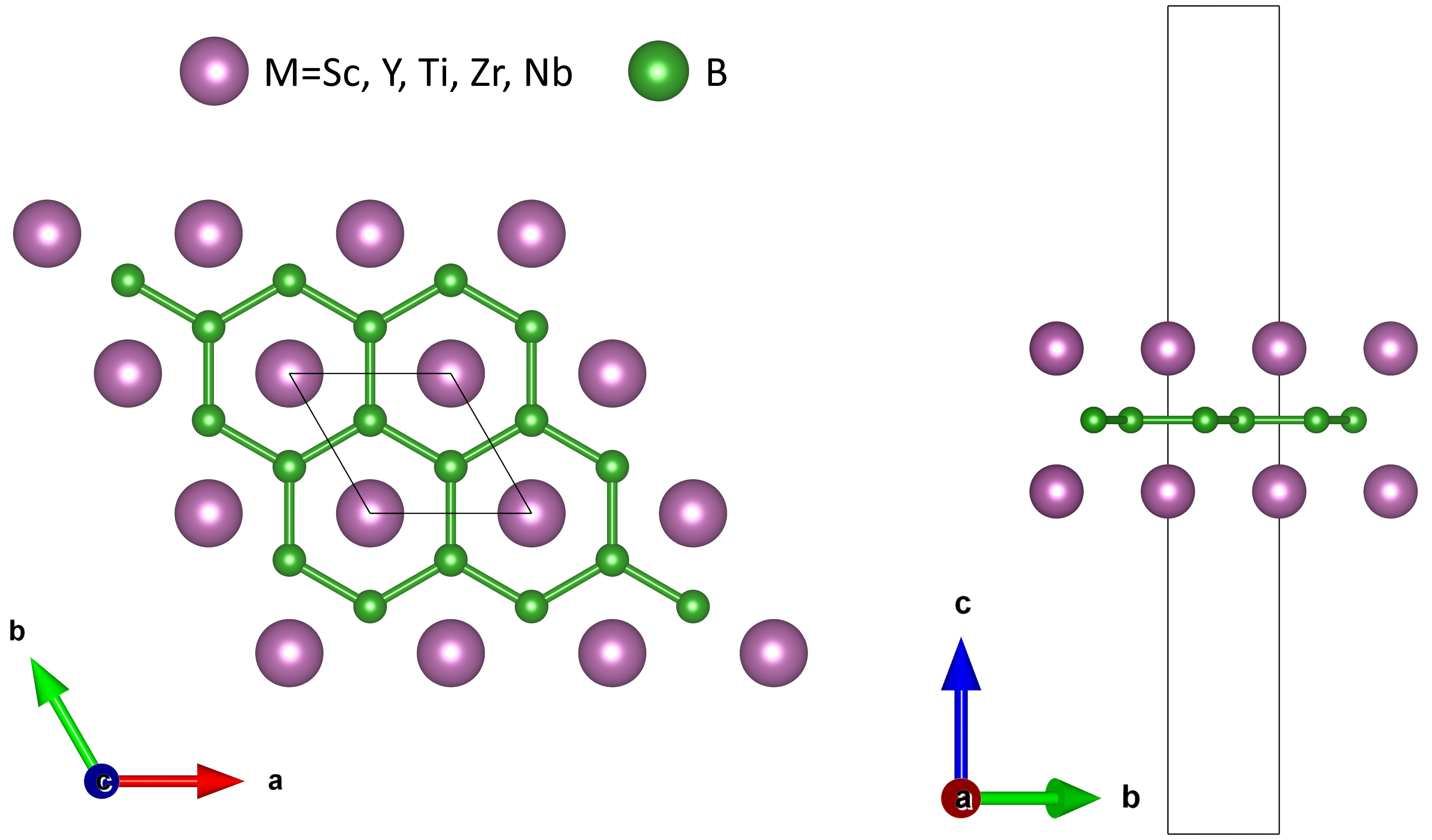}
    \caption{Top (left) and side (right) views of the optimized atomic structure of two-dimensional hexagonal 
    $\mathrm{M_2B_2}$ ($\mathrm{M = Sc, Y, Ti, Zr, Nb}$) monolayer. 
    The purple and green spheres represent the metal (M) and boron (B) atoms, respectively. The black lines indicate the primitive unit cell. 
    The structure exhibits a planar boron honeycomb layer sandwiched between two metal atom layers.}
    \label{fig:M2B2_structure}
\end{figure}

The optimized atomic structures of two-dimensional hexagonal $\mathrm{M_2B_2}$ ($\mathrm{M = Sc, Y, Ti, Zr, Nb}$) monolayers are illustrated in Figure~\ref{fig:M2B2_structure}. Each structure adopts a hexagonal primitive cell (Bravais lattice type: hP) with the extended Bravais symbol hP2 and belongs to the space group $P6/mmm$ (No.~191), indicating a highly symmetric planar configuration with inversion symmetry. In the optimized geometry within the $xy$ plane, two boron atoms occupy the fractional coordinates (0.3333, 0.6667) and (0.6667, 0.3333), forming a characteristic honeycomb network, while the metal atom is located at (0.0000, 0.0000), residing at the center of the boron hexagon. The calculated lattice constants are summarized in Table~\ref{tab:lattice_M2B2}. 

We also examined the magnetic characteristics of the hexagonal M$_2$B$_2$ (M = Sc, Y, Ti, Zr, Nb) BXene monolayers by initializing the systems in several distinct spin configurations. First, a ferromagnetic (FM) state was imposed by assigning parallel magnetic moments to all metal atoms. In addition, two antiferromagnetic (AFM) arrangements were explored: a G-type pattern, where each metal site is set to oppose the spin of all its nearest neighbors, and an A-type pattern, in which the spins within the lower M layer are aligned while those in the upper layer are reversed. Independent of the initial spin ordering, the hexagonal M$_2$B$_2$ BXene monolayers consistently relax into a metallic ground state.

\begin{table}[h!]
\centering
\caption{Optimized lattice constants $a$ and metal–metal layer thickness 
$h_{\mathrm{M}}$ of the hexagonal $\mathrm{M_2B_2}$ monolayers.}
\begin{tabular}{|c|c|c|}
\hline
M$_2$B$_2$ monolayer & $a$ (\AA) & $h_{\mathrm{M}}$ (\AA) \\
\hline
Sc$_2$B$_2$ & 3.12 & 3.46 \\
Y$_2$B$_2$  & 3.29 & 3.76 \\
Ti$_2$B$_2$ & 3.00 & 3.12 \\
Zr$_2$B$_2$ & 3.16 & 3.38 \\
Nb$_2$B$_2$ & 3.10 & 3.05 \\
\hline
\end{tabular}
\label{tab:lattice_M2B2}
\end{table}


The electronic structure of the hexagonal M$2$B$2$ monolayers is metallic, with states near the Fermi level predominantly derived from the transition-metal $d$-orbital manifold. At the $\Gamma$ point, these $d$ states split into well-defined symmetry groups: the $A'$ state originating from the $d{z^{2}}$ orbital, the $E'$ states from the in-plane $d{xy}$ and $d_{x^{2}-y^{2}}$ orbitals, and the $E''$ states from the out-of-plane $d_{yz}$ and $d_{xz}$ orbitals. The substantial $d$-orbital contribution at the Fermi level indicates a high density of available surface electrons, which can strongly influence adsorbate–surface interactions. This electronic configuration is therefore expected to facilitate charge transfer and enhance the activation of CO$_2$ upon adsorption.

\begin{figure}[h!]
    \centering
    \includegraphics[width=9cm]{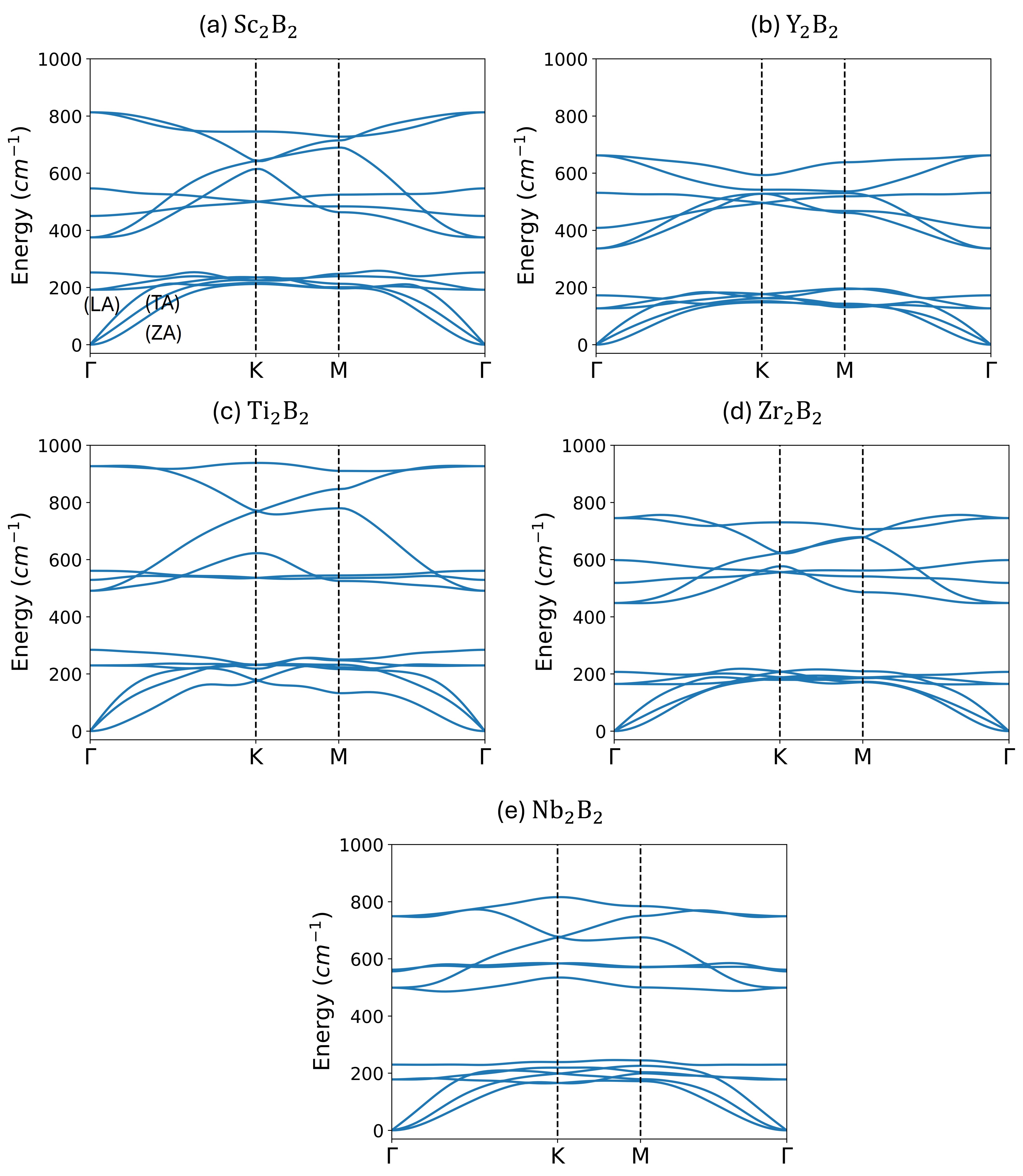}
    \caption{Phonon dispersion relations of the pristine two-dimensional 
    $\mathrm{M_2B_2}$ ($\mathrm{M = Sc, Y, Ti, Zr, Nb}$) monolayers along the 
    high-symmetry path $\Gamma$--K--M--$\Gamma$ in the Brillouin zone. 
    Panels (a)–(e) correspond to $\mathrm{Sc_2B_2}$, $\mathrm{Y_2B_2}$, 
    $\mathrm{Ti_2B_2}$, $\mathrm{Zr_2B_2}$, and $\mathrm{Nb_2B_2}$, respectively. 
    The absence of imaginary phonon modes confirms the dynamic stability of all
    $\mathrm{M_2B_2}$ monolayers.}
    \label{fig:phonon_M2B2}
\end{figure}

\begin{figure}[h!]
\centering
\includegraphics[width=6cm]{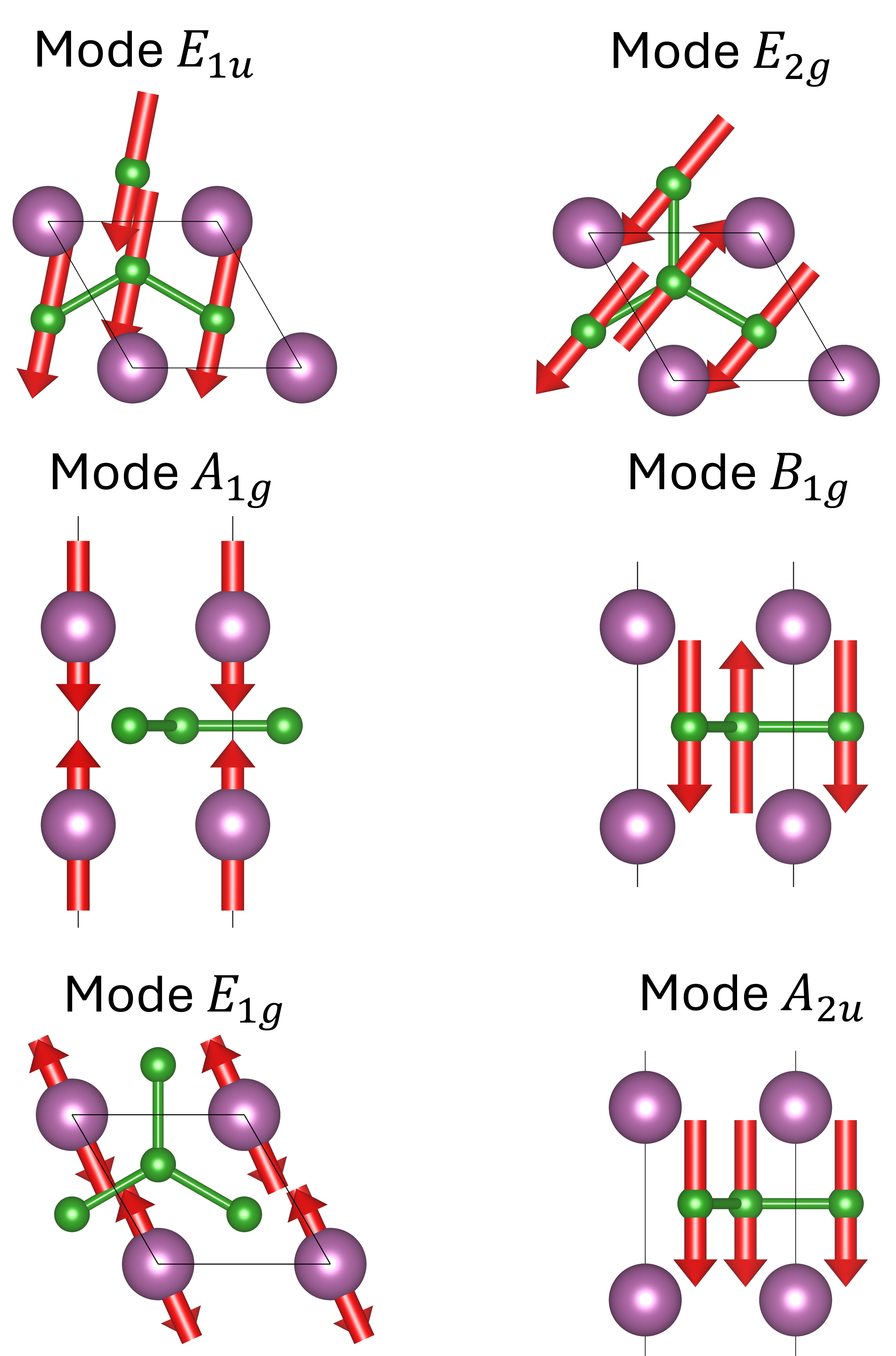}
\caption{Schematic representation of the zone-center vibrational modes of the hexagonal $\mathrm{M_2B_2}$ monolayers, illustrating the atomic displacement patterns for the symmetry modes $E_{1u}$, $E_{2g}$, $A_{1g}$, $B_{1g}$, $E_{1g}$, and $A_{2u}$. Red arrows indicate the direction and relative magnitude of atomic motions for metal (purple) and boron (green) atoms within each irreducible representation. These modes correspond to the phonon eigenvectors at the $\Gamma$ point and are used to analyze the vibrational and dynamical stability of the monolayers.}
\label{fig:phonon_modes}
\end{figure}

The dynamical stability of the two-dimensional $\mathrm{M_2B_2}$ ($\mathrm{M = Sc, Y, Ti, Zr, Nb}$) monolayers was examined by calculating their phonon dispersion relations using density functional perturbation theory. The phonon spectra for all systems show no imaginary (negative) frequencies throughout the entire Brillouin zone as shown in Figure~\ref{fig:phonon_M2B2}, confirming that each $\mathrm{M_2B_2}$ structure is dynamically stable which corresponds to previous studies~\cite{yuan2019monolayer,he2021computational,gao2021two,bo2018hexagonal,lu2024functionalizing,dai2025theoretical,han2023theoretical,han2023high,seeyangnok2025high_npj2d}. In particular, the acoustic modes near the $\Gamma$ point exhibit the expected quadratic behavior for the out-of-plane (ZA) mode and linear dispersions for the in-plane (LA and TA) modes, characteristic of stable two-dimensional materials.

\begin{table}[h!]
\centering
\caption{Selected optical phonon vibrational modes (in cm$^{-1}$) 
and their symmetry representations for pristine $\mathrm{M_2B_2}$ 
($\mathrm{M = Sc, Y, Ti, Zr, Nb}$) monolayers. 
Raman-active (R) and infrared-active (I) modes are indicated.}
\begin{tabular}{|c|c|c|c|c|c|}
\hline
Modes & Sc$_2$B$_2$ & Y$_2$B$_2$ & Ti$_2$B$_2$ & Zr$_2$B$_2$ & Nb$_2$B$_2$ \\
\hline
E$_{1g}$ (R)   & 191.4 & 126.6 & 230.2 & 164.8 & 178.1 \\
A$_{1g}$ (R)   & 252.5 & 172.5 & 284.4 & 207.1 & 229.9 \\
E$_{1u}$ (I)   & 365.7 & 339.5 & 500.3 & 440.1 & 490.9 \\
A$_{2u}$ (I)   & 454.5 & 414.0 & 535.9 & 523.7 & 565.5 \\
B$_{1g}$       & 546.6 & 531.2 & 560.7 & 598.2 & 561.9 \\
E$_{2g}$ (R)   & 812.8 & 662.2 & 926.7 & 745.1 & 749.0 \\
\hline
\end{tabular}
\label{tab:phonon_M2B2}
\end{table}

To further investigate the lattice dynamical behavior, the phonon eigenvalues and eigenvectors at the $\Gamma$ point were extracted and are summarized in Table~\ref{tab:phonon_M2B2} and Figure~\ref{fig:phonon_modes}, respectively. The highest-frequency optical modes originate predominantly from vibrations of the boron sublattice (E${1u}$, A${2u}$, B${1g}$, and E${2g}$), spanning a range of 365.7–926.7~cm$^{-1}$. In contrast, the transition-metal atoms dominate the low-frequency region, corresponding to the E${1g}$ and A${1g}$ modes, which lie between 164.8 and 284.4~cm$^{-1}$ across the series of transition metals. The high-frequency boron-related modes are a direct consequence of the strong B–B bonding network, suggested by earlier findings~\cite{seeyangnok2025high_npj2d}.

\subsection{Adsorption energetics and surface reactivity}
\begin{figure}[h!]
    \centering
    \includegraphics[width=16.5cm]{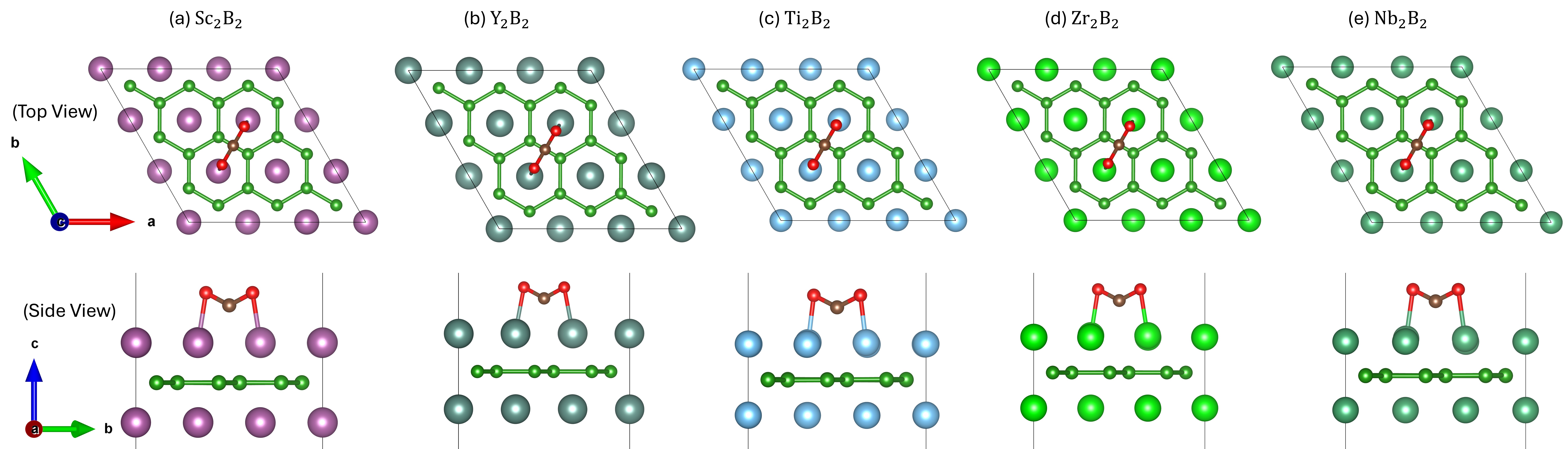}
    \caption{Top (upper panels) and side (lower panels) views of the optimized geometries of CO$_2$ adsorption on two-dimensional $\mathrm{M_2B_2}$ 
    ($\mathrm{M = Sc, Y, Ti, Zr, Nb}$) monolayers: (a) $\mathrm{Sc_2B_2}$, (b) $\mathrm{Y_2B_2}$, (c) $\mathrm{Ti_2B_2}$, (d) $\mathrm{Zr_2B_2}$, and (e) $\mathrm{Nb_2B_2}$. The red, brown, green, and colored (purple, dark green, blue, light green, and cyan) spheres represent oxygen, carbon, boron, and metal (M) atoms, respectively. The adsorption configurations show that CO$_2$ molecules are chemisorbed on the $\mathrm{M_2B_2}$ surfaces with different binding orientations, depending on the type of transition metal.}
    \label{fig:M2B2_CO2_adsorption}
\end{figure}

\begin{table*}[h!]
\centering
\caption{Combined structural, energetic, and electronic characteristics of CO$_2^{\delta-}$ adsorption on $\mathrm{M_2B_2}$ monolayers. Adsorption energies $E_{\mathrm{ads}}$ are in eV; C--O bond lengths $d_{\mathrm{CO}}$ are in \AA; O--C--O angles are in degrees; lpCOHP$(E_f)$ is in eV; and charge transfer (CT) values are in units of $e$ from Löwdin and Bader analyses.}
\begin{tabular}{|c|c|c|c|c|c|c|}
\hline
System & $E_{\mathrm{ads}}$ (eV) & $d_{\mathrm{CO}}$ (\AA) & O--C--O ($^\circ$) 
& lpCOHP$(E_f)$ (eV) & Löwdin CT ($e$) & Bader CT ($e$) \\
\hline
CO$_2$ (isolated) & -- & 1.17 & 180.00 & -18.29 & -- & -- \\
\hline
Sc$_2$B$_2$ & -2.15 & 1.28 & 129.97 & -13.98 & 0.55 & 1.43 \\
Y$_2$B$_2$  & -1.93 & 1.28 & 129.05 & -13.91 & 0.20 & 1.46 \\
Ti$_2$B$_2$ & -2.16 & 1.29 & 131.96 & -14.16 & 0.52 & 1.34 \\
Zr$_2$B$_2$ & -1.87 & 1.28 & 131.05 & -14.20 & 0.13 & 1.35 \\
Nb$_2$B$_2$ & -1.84 & 1.27 & 132.14 & -14.43 & 0.09 & 1.17 \\
\hline
\end{tabular}
\label{tab:CO2_combined_properties}
\end{table*}
Figure~\ref{fig:M2B2_CO2_adsorption} presents the optimized adsorption configurations of CO$_2$ molecules on the surfaces of two-dimensional $\mathrm{M_2B_2}$ ($\mathrm{M = Sc, Y, Ti, Zr, Nb}$) monolayers, shown from both top and side views. In all cases, the CO$_2$ molecule interacts directly with the metal site, indicating that the transition-metal atoms. The CO$_2$ molecule undergoes noticeable bending upon adsorption, suggesting partial activation due to charge transfer between the molecule and the substrate. This behavior demonstrates the chemisorptive nature of the interaction rather than weak physisorption. The bending angle of CO$_2$ are particularly sensitive to the metal type, with lighter elements (e.g., Sc and Ti) exhibiting larger molecular distortion compared to heavier elements (e.g., Y, Zr, and Nb).

As summarized in Table~\ref{tab:CO2_combined_properties}, the adsorption energies of CO$_2$ on the $\mathrm{M_2B_2}$ ($\mathrm{M = Sc, Y, Ti, Zr, Nb}$) monolayers reveal strong chemisorption across all systems, with $E_{\mathrm{ads}}$ values ranging from $-1.84$ to $-2.16$~eV. Among the studied materials, Ti$_2$B$_2$ and Sc$_2$B$_2$ exhibit the strongest adsorption, with $E_{\mathrm{ads}} = -2.16$~eV and $-2.15$~eV, respectively, while Nb$_2$B$_2$ and Zr$_2$B$_2$ show slightly weaker but still substantial binding strengths. Concurrently, the CO$_2$ molecule undergoes significant structural deformation upon adsorption when compared to its isolated configuration. In the gas phase, CO$_2$ has a C--O bond length of 1.17~\AA{} and a linear O--C--O angle of $180^\circ$; however, upon adsorption, the C--O bonds elongate to 1.27--1.29~\AA{}, and the O--C--O angle decreases markedly to approximately $129^\circ$--$132^\circ$. This pronounced molecular bending and bond stretching indicate weakening of the C--O bonds and reflect substantial activation of CO$_2$ on the $\mathrm{M_2B_2}$ surfaces. These structural and energetic trends confirm the chemisorptive nature of the interaction and demonstrate that $\mathrm{M_2B_2}$ monolayers, particularly those containing early transition metals, provide highly reactive platforms capable of promoting CO$_2$ activation.

\subsection{Charge Transfer and Bonding Characteristics}
The Löwdin charge analysis~\cite{lowdin1950non} further confirms that CO$_2$ receives a substantial amount of electronic charge upon adsorption on the $\mathrm{M_2B_2}$ monolayers. As shown in Table~\ref{tab:CO2_combined_properties}, the total electronic population on the CO$_2$ molecule increases from 16 electrons in the isolated state to values between approximately 16.09 and 16.55 electrons after adsorption. This corresponds to charge transfer values of 0.55$e$, 0.20$e$, 0.52$e$, 0.13$e$, and 0.09$e$ for the Sc-, Y-, Ti-, Zr-, and Nb-based systems, respectively. The relatively large charge transfer observed for Sc$_2$B$_2$ and Ti$_2$B$_2$ correlates with their stronger adsorption energies and the more pronounced structural deformation of the CO$_2$ molecule, indicating a higher degree of activation. In contrast, the smaller charge transfer in Zr$_2$B$_2$ and Nb$_2$B$_2$ reflects weaker interaction strength and reduced bending of CO$_2$. 

The Bader charge analysis~\cite{tang2009grid} as shown in Table~\ref{tab:CO2_combined_properties} also supports this trend, revealing even larger electron transfer to the adsorbed CO$_2$ molecule. The computed charge transfer values are 1.46$e$, 1.43$e$, 1.34$e$, 1.35$e$, and 1.17$e$ for the Sc-, Y-, Ti-, Zr-, and Nb-based systems, respectively, indicating that CO$_2$ is reduced to a CO$_2^{\delta-}$ species on all $\mathrm{M_2B_2}$ surfaces. Although the absolute values differ from those of the Löwdin scheme due to the distinct charge-partitioning methods, both analyses consistently exhibit the same qualitative trend: early transition-metal borides (Sc and Y) transfer more charge to CO$_2$, whereas Nb donates the least. The larger Bader charges in Sc$_2$B$_2$ and Y$_2$B$_2$ correlate with more severe molecular deformation and stronger activation, while the smaller charge in Nb$_2$B$_2$ correlates with weaker distortion and reduced reactivity. Together, the Löwdin and Bader results provide compelling evidence that electron donation from the $\mathrm{M_2B_2}$ surface is the primary driving force for weakening the internal C--O bonds and promoting CO$_2$ activation.

\begin{figure}[h!]
\centering
\includegraphics[width=7.5cm]{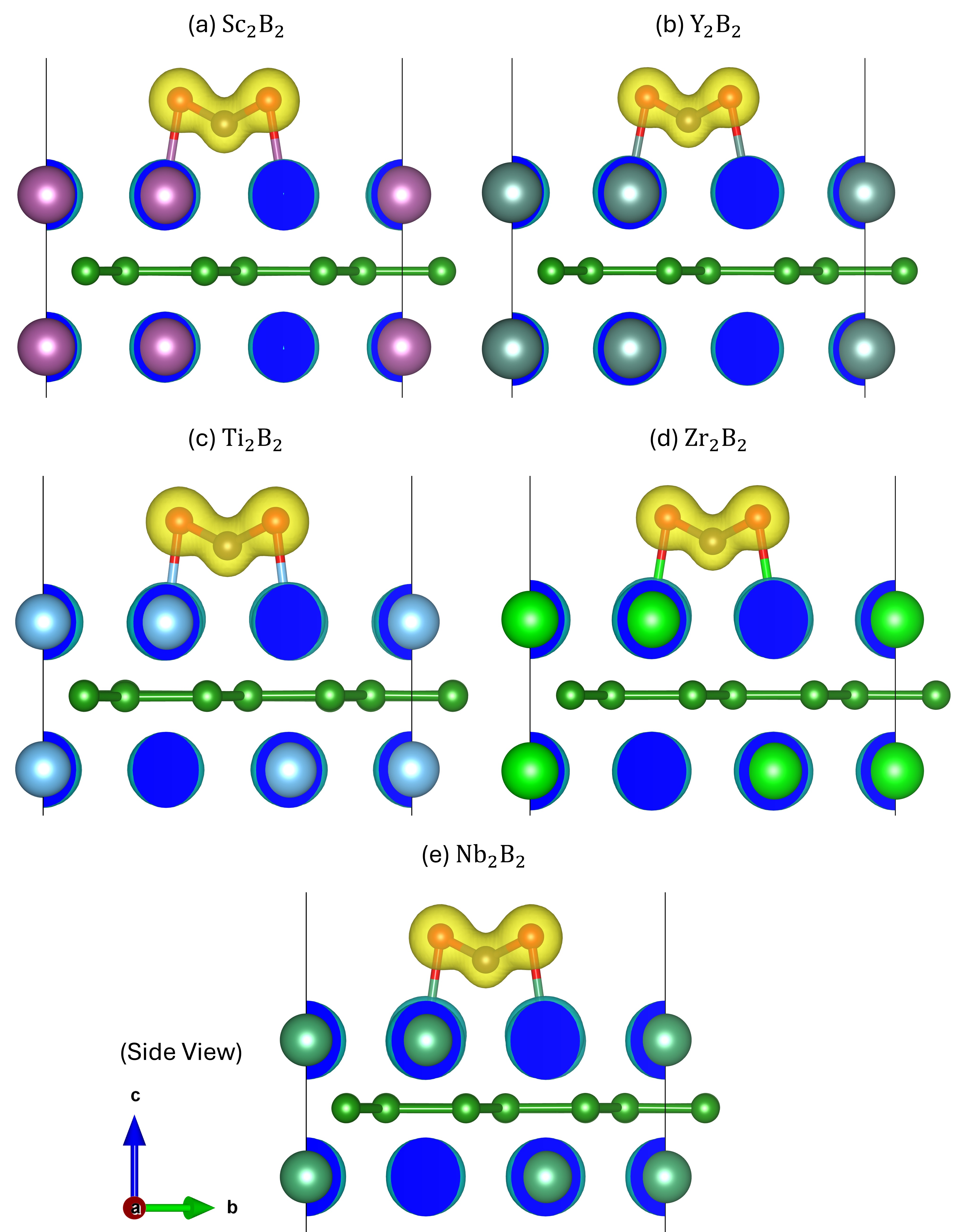}
\caption{Charge density difference plots for CO$_2$ adsorption on the surface at an isosurface level of 0.15. Yellow and cyan denote charge accumulation and depletion, respectively, indicating interfacial charge transfer upon adsorption.}
\label{fig:charge-difference}
\end{figure}

To further confirm the presence of charge transfer, as suggested by the Löwdin and Bader charge analyses, we analyzed the charge density difference, defined as
\begin{equation}
\Delta \rho = \rho_{\mathrm{M_2B_2 + CO_2}} - \rho_{\mathrm{M_2B_2}} - \rho_{\mathrm{CO_2}},
\end{equation}
where $\rho_{\mathrm{total}}$ denotes the charge density of the combined adsorption system, while $\rho_{\mathrm{surface}}$ and $\rho_{\mathrm{CO_2}}$ represent the charge densities of the isolated surface and the CO$_2$ molecule, respectively. The charge density difference $\Delta \rho$ confirms charge transfer from the M$_2$B$_2$ surface to the CO$_2$ molecule, as shown in Fig.~\ref{fig:charge-difference}. This is evidenced by charge accumulation (yellow) on the CO$_2$ molecule and charge depletion (cyan) on the M$_2$B$_2$ surface corresponding to the Löwdin and Bader charge analyses.

\begin{figure}[h!]
    \centering
    \includegraphics[width=8cm]{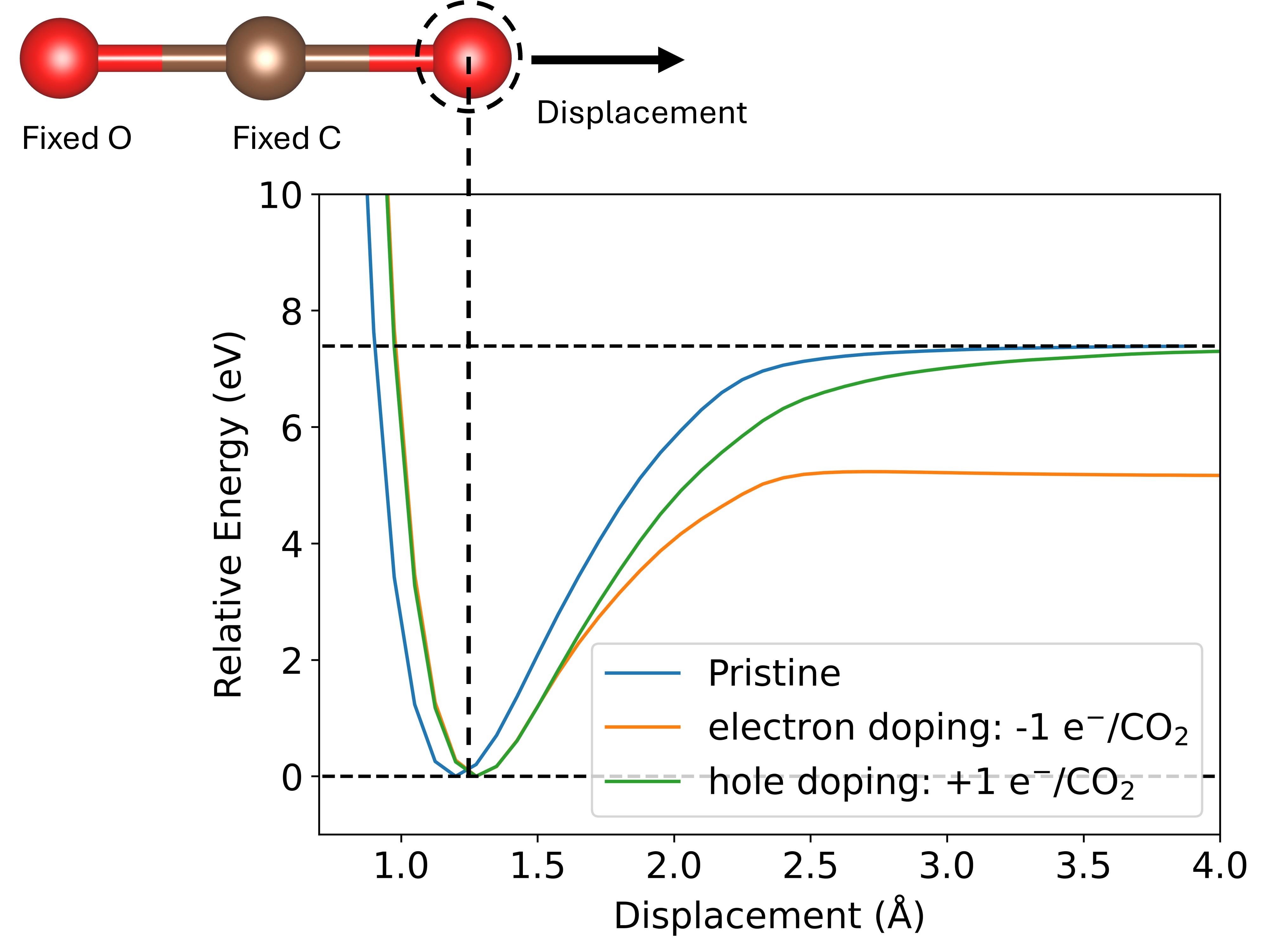}
    \caption{Schematic representation (top) of the constrained desorption pathway, where one oxygen atom and the carbon atom of CO$_2$ are fixed while the second oxygen atom is displaced away from the adsorption site. The corresponding relative energy profiles (bottom) are shown for pristine, electron-doped ($-1\,e^{-}$/CO$_2$), and hole-doped ($+1\,e^{-}$/CO$_2$) $\mathrm{M_2B_2}$ systems as a function of the displacement of the free oxygen atom. Electron doping significantly lowers the desorption energy barrier, making CO$_2$ easier to detach, whereas hole doping slightly lowers the barrier relative to the pristine system.}
    \label{fig:CO2_desorption_energy_curve}
\end{figure}

The influence of charge transfer on CO$_2$ bond strength is further reflected in the desorption energy profiles shown in Figure~\ref{fig:CO2_desorption_energy_curve}. Both electron and hole doping reduce the depth of the potential well relative to pristine CO$_2$, but to markedly different degrees. The electron-doped case (–1 e$^{-}$/CO$_2$) exhibits the shallowest potential well, indicating that the additional electronic charge most effectively weakens the internal C–O bonds. Hole doping (+1 e$^{-}$/CO$_2$) also produces a slightly shallower well than the pristine molecule, though the effect is considerably weaker. This behavior is fully consistent with the adsorption trends observed for the $\mathrm{M_2B_2}$ monolayers: systems such as Sc$_2$B$_2$ and Ti$_2$B$_2$, which transfer the largest amount of charge to the adsorbed molecule, induce the strongest CO$_2$ activation—manifested by pronounced C–O bond elongation, significant bending, and more adsorption energies. In contrast, monolayers that donate less charge, such as Zr$_2$B$_2$ and Nb$_2$B$_2$, produce weaker distortions and reduced activation. Thus, the isolated CO$_2$ desorption curves offer a clear mechanistic interpretation of the structural and energetic trends across the $\mathrm{M_2B_2}$ series: greater electron donation leads to stronger activation and more pronounced weakening of the C–O bonds.

\begin{figure}[h!]
    \centering
    \includegraphics[width=8cm]{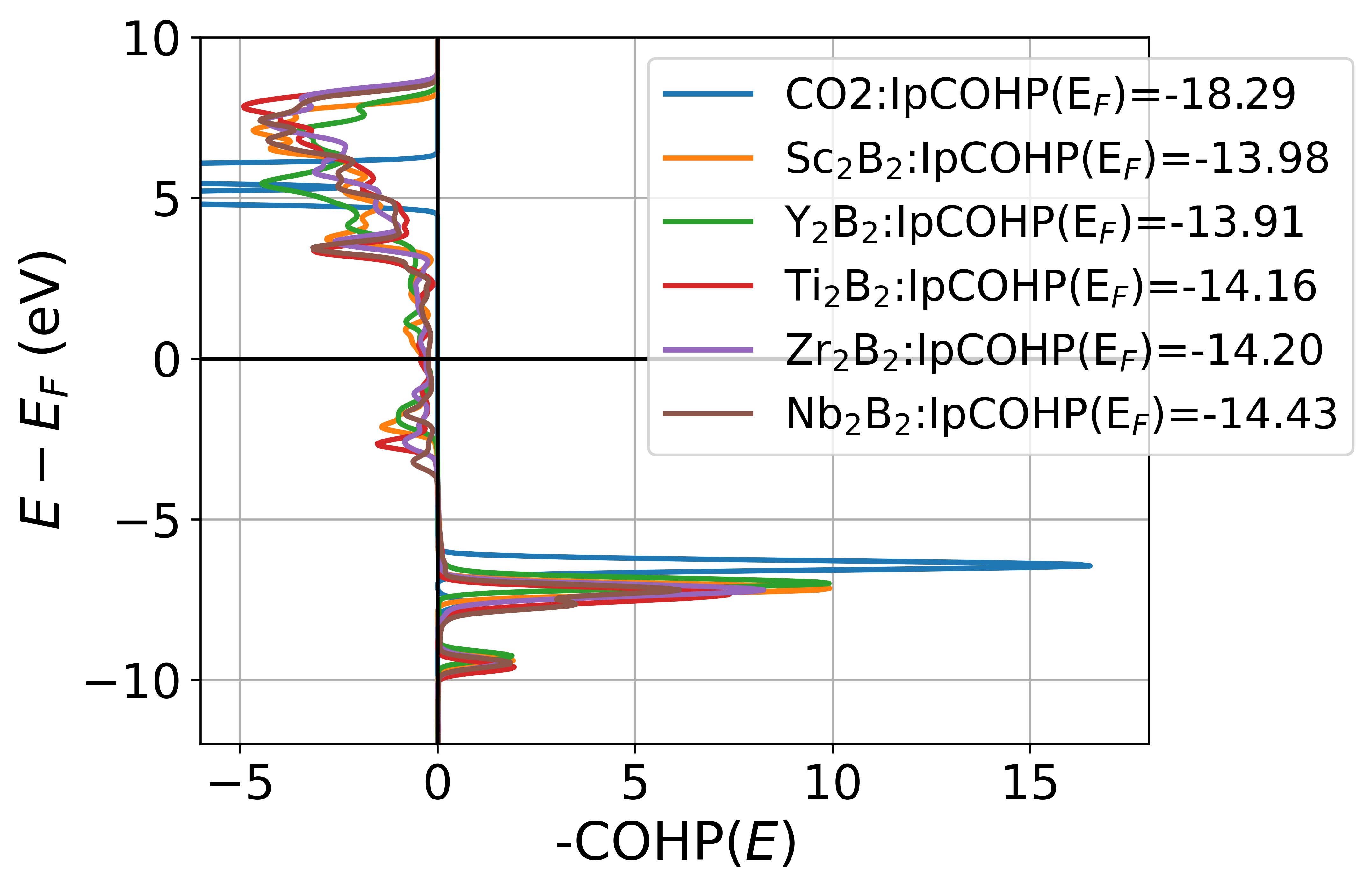}
    \caption{Crystal orbital Hamilton population (COHP) analysis of the C--O bonds in CO$_2$ for the isolated molecule and for CO$_2$ adsorbed on the $\mathrm{M_2B_2}$ ($\mathrm{M = Sc, Y, Ti, Zr, Nb}$) monolayers. The $-$COHP($E$) curves are plotted as a function of energy referenced to the Fermi level ($E - E_F$). More negative IpCOHP$(E_f)$ values correspond to stronger C--O bonding.} 
    \label{fig:CO2_COHP}
\end{figure}

\begin{figure}[h!]
\centering
\includegraphics[width=16.5cm]{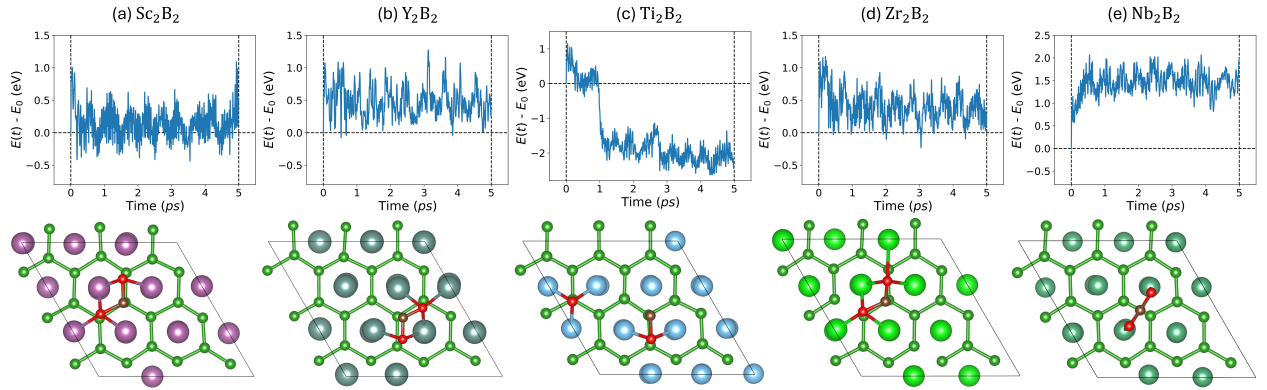}
\caption{
Ab initio molecular dynamics (AIMD) simulations of CO$_2$ adsorbed on 
(a) Sc$_2$B$_2$, (b) Y$_2$B$_2$, (c) Ti$_2$B$_2$, (d) Zr$_2$B$_2$, and (e) Nb$_2$B$_2$ monolayers at 300~K. Top panels show the time evolution of the total energy relative to the initial value, $E(t)-E_{0}$, over a 5~ps trajectory, demonstrating thermal stability and the absence of desorption events under ambient conditions. Bottom panels display the final atomic configurations obtained at the end of the AIMD runs (top view).}
\label{fig:AIMD_M2B2_CO2}
\end{figure}

The crystal orbital Hamilton population (COHP) analysis as shown in Figure~\ref{fig:CO2_COHP} provides direct insight into how adsorption on the $\mathrm{M_2B_2}$ monolayers modifies the intrinsic C–O bonding in CO$_2$. For isolated CO$_2$, the strongly negative IpCOHP$(E_f)$ value of $-18.29$~eV reflects the robust covalent character of the C–O double bonds. Upon adsorption, however, the IpCOHP$(E_f)$ values become substantially less negative, ranging from $-13.91$ to $-14.43$~eV as summarized in Table~\ref{tab:CO2_combined_properties}. This reduction indicates a pronounced weakening of the internal C–O bonds, consistent with the elongated C–O bond lengths and bent O–C–O angles obtained from structural optimization. The shift toward less negative IpCOHP$(E_f)$ values reveals increased population of antibonding $\pi^\ast$ orbitals as shown in Figure~\ref{fig:CO2_COHP} just below Fermi level, lowering the C–O bond order and transforming the molecule into a chemically activated CO$_2^{\delta-}$ species.

The trend in COHP results correlates strongly with the amount of electron transfer to CO$_2$, as revealed by the Löwdin and Bader charge analyses also listed in Table~\ref{tab:CO2_combined_properties}. Systems exhibiting large charge transfer—such as Sc$_2$B$_2$ and Ti$_2$B$_2$—show the greatest reductions in IpCOHP$(E_f)$, reflecting stronger filling of antibonding orbitals and consequently more severe weakening of the C–O bonds. In contrast, Zr$_2$B$_2$ and Nb$_2$B$_2$, which transfer significantly less charge, maintain more negative IpCOHP$(E_f)$ values and exhibit correspondingly less absorption energies. Together, the COHP and charge-transfer results provide a consistent and unified picture: electron donation from the $\mathrm{M_2B_2}$ surface is the primary mechanism responsible for weakening the C–O bonds and driving CO$_2$ activation across the examined monolayers.

\subsection{Thermal Effects on CO$_2^{\delta -}$ Adsorption}
Further investigation of thermal effects was carried out using ab initio molecular dynamics (AIMD) simulations of CO$_2$ adsorbed on hexagonal M$_2$B$_2$ monolayers at 300~K within the NVT ensemble, with the corresponding energy fluctuations and final configurations shown in Figure~\ref{fig:AIMD_M2B2_CO2}. Interestingly, the AIMD trajectories reveal multiple distinct final configurations, consistent with the structural variations also found in static optimizations performed with different initial CO$_2$ orientations (see Appendix). These final AIMD results are consistent with the optimized structures obtained from configuration (g), where the CO$_2$ molecule aligns itself along the boron honeycomb network, leading to the most stable adsorption configuration.

For Nb$_2$B$_2$, the final CO$_2$ geometry remains essentially unchanged, matching the optimized adsorption structure reported in Figure~\ref{fig:M2B2_CO2_adsorption}. In contrast, Sc$_2$B$_2$, Y$_2$B$_2$, and Zr$_2$B$_2$ exhibit slight re-orientations of the reactive CO$_2^{\delta -}$ species: although the molecule stays chemisorbed, it realigns itself with the underlying boron network during the simulation. The most pronounced thermal effect occurs for Ti$_2$B$_2$, where one O atom dissociates from the activated CO$_2^{\delta -}$, resulting in spontaneous cleavage into CO and O fragments at room temperature. These results demonstrate that the activated CO$_2^{\delta -}$ species is highly sensitive to thermal fluctuations, and that even mild thermal energy at 300~K can significantly rearrange—or fully break—the molecular configuration on the most reactive M$_2$B$_2$ surfaces.

\section{Conclusion}
In this work, we performed a comprehensive first-principles investigation of CO$_2$ adsorption and activation on hexagonal transition-metal diboride monolayers, M$_2$B$_2$ (M = Sc, Y, Ti, Zr, Nb), to evaluate their potential as two-dimensional materials for CO$_2$ capture and surface reactivity. All pristine M$_2$B$_2$ monolayers were confirmed to be mechanical, dynamically ,and energetically stable. Across the entire series, CO$_2$ adsorption was found to be strongly exothermic, with adsorption energies in the range of $-1.84$ to $-2.16$ eV (or $-1.98$ to $-4.42$), reflecting robust chemisorption rather than weak physisorption. This strong interaction induces pronounced structural deformation of the molecule, including elongation of the C–O bonds and bending of the O–C–O angle, signaling significant activation of CO$_2$ upon binding.

Charge-transfer analyses using Löwdin and Bader schemes consistently revealed substantial electron donation from the M$_2$B$_2$ surface to the adsorbed CO$_2$, yielding a negatively charged CO$_2^{\delta-}$ species. Early transition metals (Sc, Ti, Y) exhibited the largest charge transfer and correspondingly the strongest adsorption and molecular deformation. These trends were further corroborated by COHP calculations, which showed a marked decrease in IpCOHP$(E_f)$ values for the C–O bonds after adsorption. This shift toward less negative IpCOHP$(E_f)$ indicates enhanced population of antibonding $\pi^\ast$ states, leading to significant weakening of the internal C–O bonds—an essential electronic signature of CO$_2$ activation.

Importantly, ab initio molecular dynamics simulations revealed that the activated CO$_2^{\delta-}$ species is highly sensitive to thermal fluctuations. At 300~K, the M$_2$B$_2$ surfaces exhibited distinct dynamical behavior: Nb$_2$B$_2$ retained a stable chemisorbed configuration, while Sc$_2$B$_2$, Y$_2$B$_2$, and Zr$_2$B$_2$ showed moderate reorientation of the adsorbed molecule. Notably, Ti$_2$B$_2$ induced spontaneous dissociation of CO$_2^{\delta-}$ into CO and O during the simulation. These results indicate that once CO$_2$ is activated to a bent, electron-rich CO$_2^{\delta-}$ form, its weakened C–O bonds become highly susceptible to thermally driven rearrangement or even bond cleavage at room temperature. The optimization of configuration (g) as discussed in The Appendix also show that the 

Overall, the combined structural, electronic, and bonding analyses demonstrate that hexagonal M$_2$B$_2$ monolayers, particularly those incorporating early transition metals, provide highly reactive platforms capable of activating CO$_2$ through a charge-transfer-driven mechanism. These findings position two-dimensional transition-metal diborides as promising candidates for future CO$_2$ capture, activation, and potentially catalytic conversion technologies. The insights established here offer a foundation for rational design and chemical tuning of boride-based 2D materials aimed at advancing next-generation carbon management strategies.

    \section*{Data Availability}
    The data that support the findings of this study are available from the corresponding
    authors upon reasonable request.
    
    \section*{Code Availability}
    The first-principles DFT calculations were performed using the open-source Quantum ESPRESSO package, available at \url{https://www.quantum-espresso.org}, along with pseudopotentials from the Quantum ESPRESSO pseudopotential library at \url{https://pseudopotentials.quantum-espresso.org/}.
\section*{Acknowledgments}
	This research project is supported by the Second Century Fund (C2F), Chulalongkorn University. We acknowledge the supporting computing infrastructure provided by NSTDA, CU, CUAASC, NSRF via PMUB [B05F650021, B37G660013] (Thailand).

    \section*{Author Contributions}
    Jakkapat Seeyangnok conceptualized the research idea, performed all calculations and visualizations, analyzed the results, and wrote the manuscript. Rungkiat Nganglumpoon and Joongjai Panpranot contributed to the analysis and manuscript writing. Udomsilp Pinsook coordinated the project, contributed to the analysis, and edited the final manuscript.

    \section*{Conflict  of Interests}
    The authors declare no competing financial or non-financial interests.

\section*{Appendix}
\subsection*{Appendix A: Configurational CO$_2$ substitutions}
\begin{figure}[ht]
\centering
\includegraphics[width=12cm]{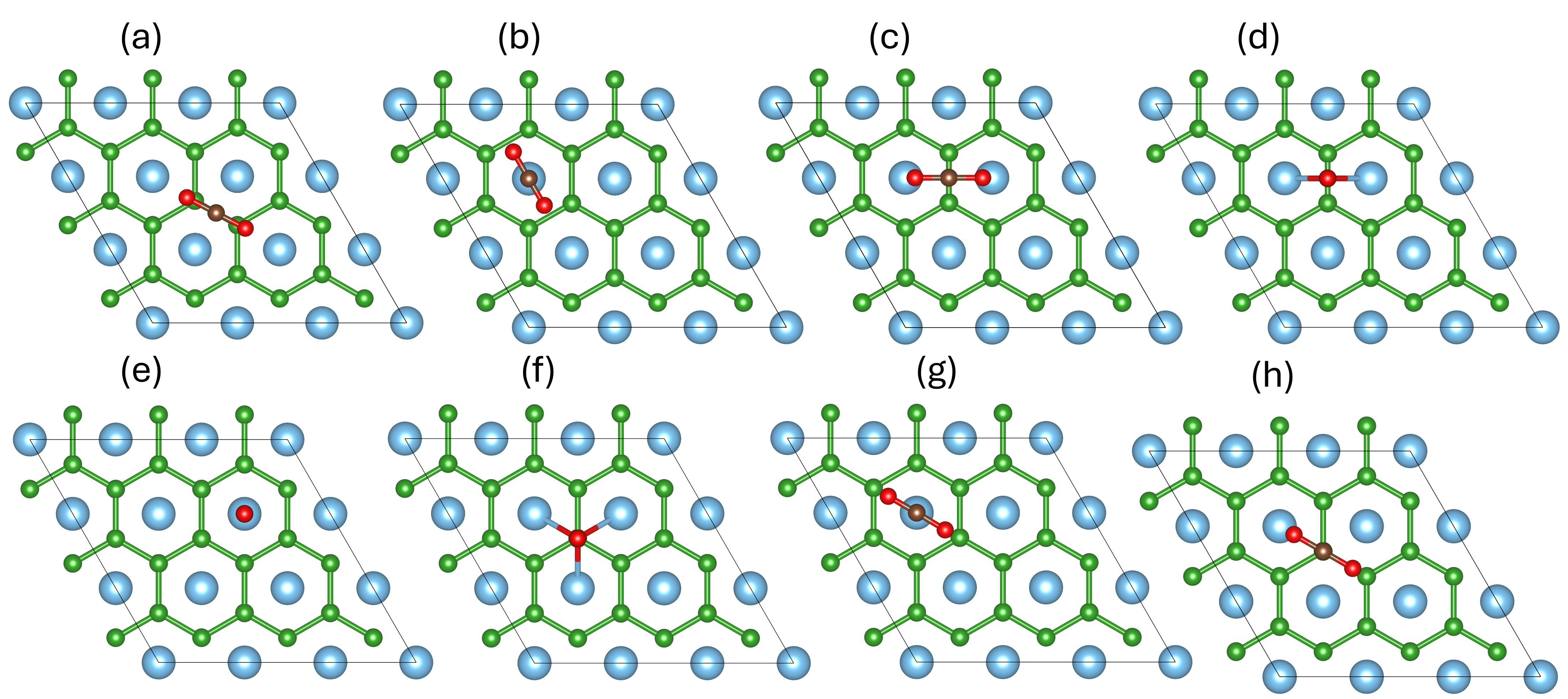}
\caption{
Top-view structures of the considered initial CO$_2$ adsorption configurations on the hexagonal M$_2$B$_2$ monolayer. (a)–(c) and (g)–(h): CO$_2$ aligned parallel to the M$_2$B$_2$ surface at different adsorption sites. (d)–(f): CO$_2$ aligned vertically relative to the monolayer surface. Green, blue, brown, and red spheres represent B, transition-metal, C, and O atoms, respectively.}
\label{fig:CO2-configurations}
\end{figure}

To explore the adsorption behavior of CO$_2$ molecules on the hexagonal M$_2$B$_2$ monolayer, various initial adsorption configurations were constructed, as illustrated in Figure~\ref{fig:CO2-configurations}. These configurations include different adsorption sites (top, bridge, and hollow positions) and molecular orientations. In configurations (d), (e), and (f), the CO$_2$ molecule is oriented vertically with respect to the M$_2$B$_2$ surface, whereas in configurations (a)–(c) and (g)–(h), CO$_2$ is aligned parallel to the monolayer.

For the vertical adsorption configurations (d), (e), and (f), the CO$_2$ molecules remain almost upright after structural relaxation, similar to their initial configuration, indicating weak physisorption with relatively small adsorption energies, -0.14 to -1.24~eV, as summarized in Table~\ref{tab:vertical-adsorption}.

\begin{table}[ht]
\centering
\caption{Adsorption energies (in eV) of CO$_2$ for configurations (b), (d), (e), (f), (g) and (h) on M$_2$B$_2$ monolayers.}
\label{tab:vertical-adsorption}
\begin{tabular}{lccccc}
\hline
Configuration & Sc & Y & Ti & Zr & Nb \\
\hline
(a,c) & -2.15 & -1.93 & -2.16 & -1.87 & -1.84 \\
(b) & -1.24 & -0.98 & -1.05 & -0.60 & -0.14 \\
(d) & -0.10 & -0.13 & -0.24 & -0.09 & -0.10 \\
(e) & -0.21 & -1.44 & -0.22 & -0.20 & -0.26 \\
(f) & -0.08 & -2.06 & -0.26 & -0.08 & -0.10 \\
(g) & -3.38 & -2.22 & -4.42 &  -4.93 & -1.98 \\
(h) & -2.26 & -2.06 & -2.57 & -2.15 & -3.75 \\
\hline
\end{tabular}
\end{table}

\begin{figure}[ht]
\centering
\includegraphics[width=6cm]{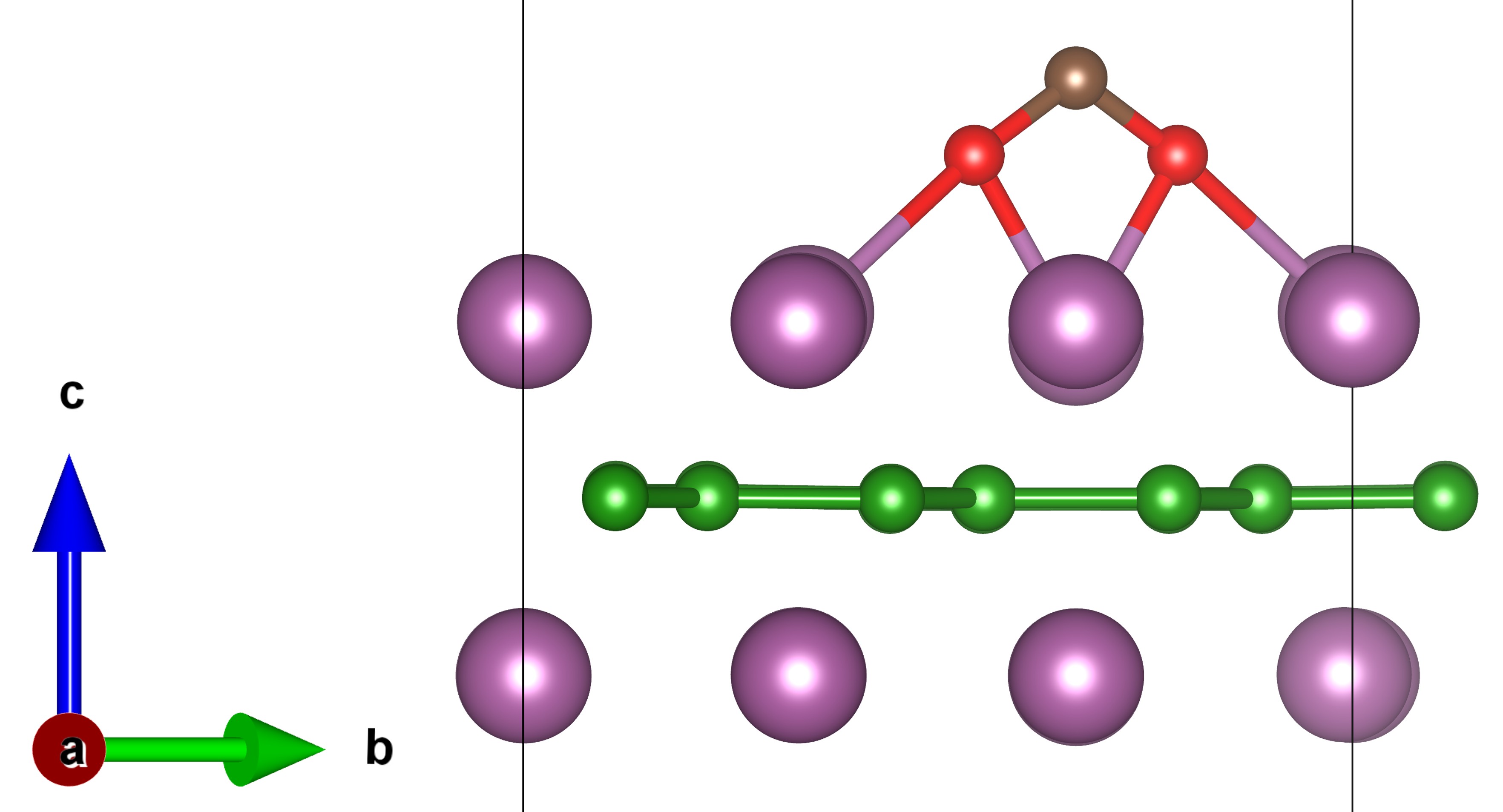}
\caption{Side-view structure of the CO$_2$ adsorption configuration (b) on the M$_2$B$_2$ monolayer. After relaxation, the CO$_2$ molecule becomes bent due to the interaction with the surface. Brown, red, green, and purple spheres represent C, O, B, and transition-metal atoms, respectively.}
\label{fig:CO2-b-case}
\end{figure}

The initial configurations (a) and (c) converge to the same optimized adsorption structure, as shown in Figure~\ref{fig:M2B2_CO2_adsorption}, which has already been discussed in the main text. Configuration (b) also results in a bent CO$_2$ geometry after relaxation (Figure~\ref{fig:CO2-b-case}), indicating partial activation of the molecule. However, as summarized in Table~\ref{tab:vertical-adsorption}, the corresponding adsorption energies of configuration (b) are generally smaller than those of configurations (a) and (c), revealing a weaker interaction with the M$_2$B$_2$ surface.

\begin{figure}[ht]
\centering
\includegraphics[width=10cm]{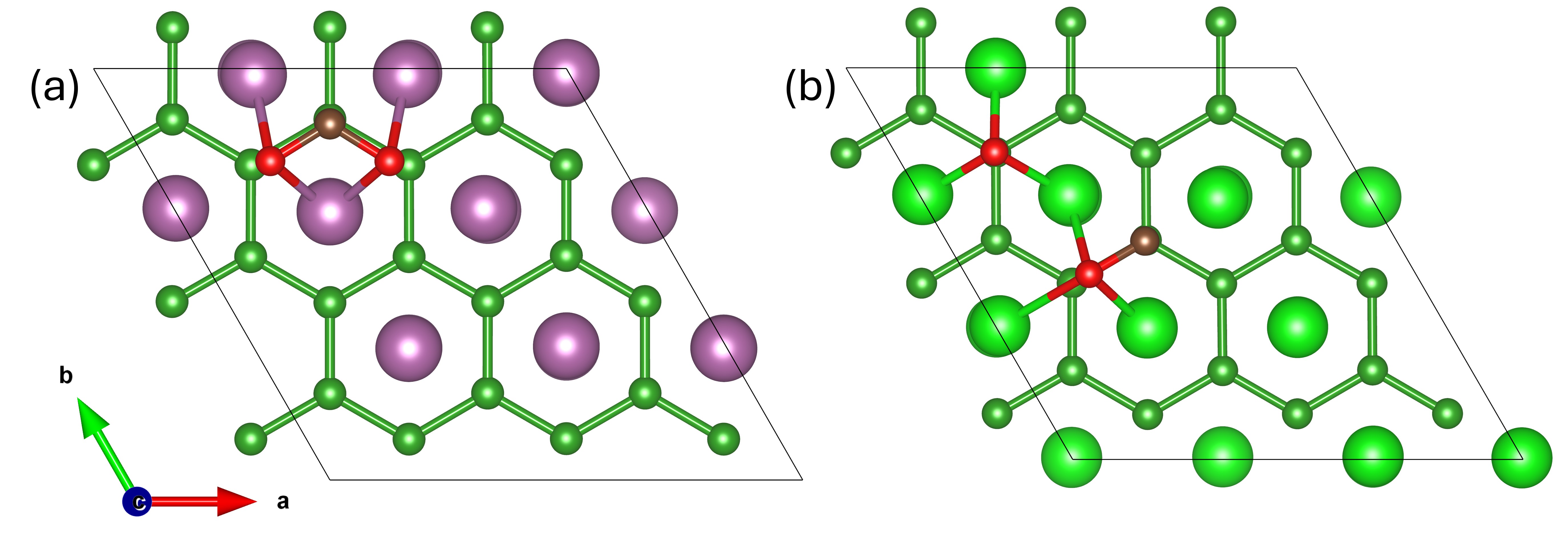}
\caption{Final optimized adsorption configurations obtained from the initial (g) configuration of CO$_2$ on M$_2$B$_2$ monolayers. Panel (a) shows that the structure converges to the same adsorption geometry for Sc, Y, Ti, and Nb systems. Panel (b) corresponds to the Zr$_2$B$_2$ case, where one O atom dissociates from the CO$_2$ molecule, indicating strong chemisorption and partial CO$_2$ decomposition. Brown, red, green, and purple spheres represent C, O, B, and transition-metal atoms, respectively.}
\label{fig:g-to-final}
\end{figure}

Finally, the initial configurations (g) and (h) lead to very interesting outcomes. For some transition metals, the optimized structures show a dissociation of one O atom from the CO$_2$ molecule, consistent with the final AIMD result shown in Figure~\ref{fig:AIMD_M2B2_CO2}(c) for Ti$_2$B$_2$. For configuration (g), most systems relax into a similar adsorption geometry to the final MD state, where the CO$_2$ molecule aligns itself along the boron honeycomb network, as illustrated in Figure~\ref{fig:AIMD_M2B2_CO2}(a,b,d,e) and Figure~\ref{fig:g-to-final}(a). These optimized configurations yield even lower adsorption energies ranging from -19.8 to -4.42~eV, as summarized in Table~\ref{tab:vertical-adsorption}, in good agreement with our AIMD results. Interestingly, in the case of Zr$_2$B$_2$, one O atom detaches from CO$_2$, indicating a strong chemisorption interaction and partial CO$_2$ decomposition, as shown in Figure~\ref{fig:g-to-final}.

\begin{figure}[ht]
\centering
\includegraphics[width=10cm]{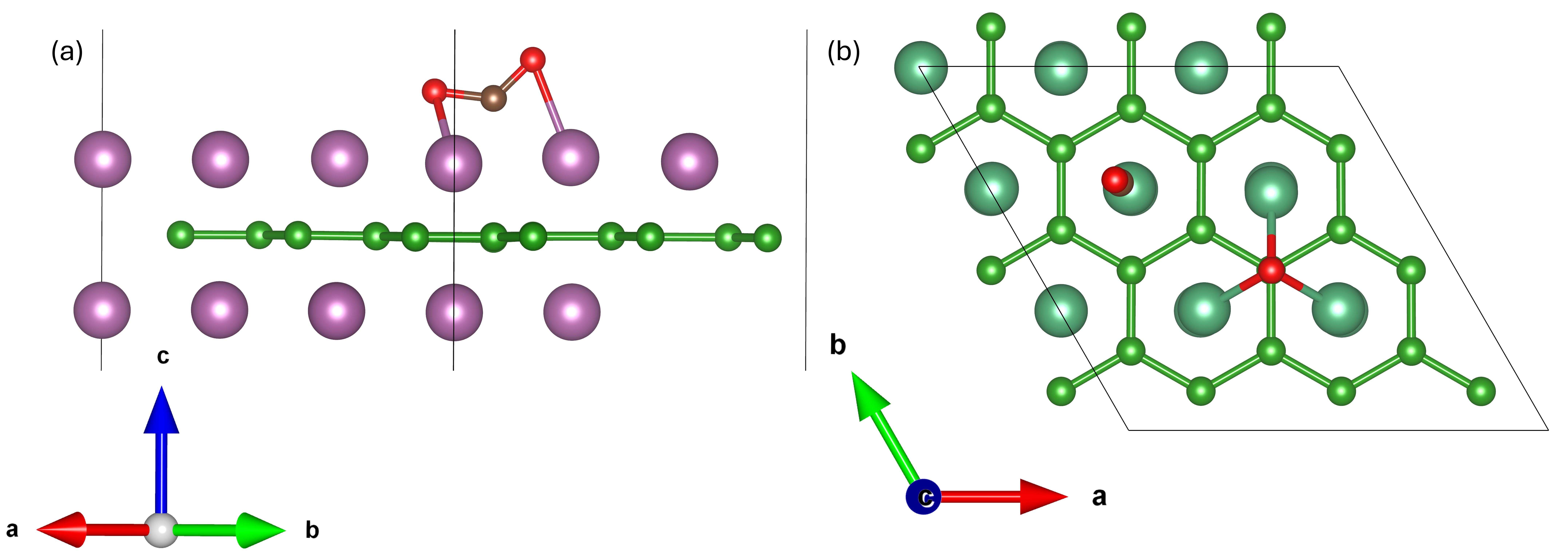}
\caption{
Final optimized adsorption configurations obtained from the initial (h) configuration of CO$_2$ on M$_2$B$_2$ monolayers. 
(a) Most systems converge to a slightly tilted adsorption geometry similar to those observed in configurations (a) and (b). 
(b) For Nb$_2$B$_2$, one O atom dissociates from the CO$_2$ molecule, indicating a strong chemisorption interaction and partial CO$_2$ decomposition. 
Brown, red, green, and the larger spheres represent C, O, B, and M atoms, respectively. 
The lattice vectors (\textbf{a}, \textbf{b}, \textbf{c}) are shown for reference.
}
\label{fig:h-to-final}
\end{figure}

For configuration (h), the optimized structure is similar to the final configurations obtained from (a) and (b), but with a slightly tilted CO$_2$ orientation, as shown in Figure~\ref{fig:h-to-final}(a). 
Interestingly, in the case of Nb$_2$B$_2$, one O atom dissociates from CO$_2$, indicating strong chemisorption and partial CO$_2$ decomposition, as presented in Figure~\ref{fig:h-to-final}(b).

\bibliographystyle{unsrt}
\bibliography{references}

\end{document}